\documentclass[vecphys]{svmult}

\usepackage{makeidx}         
\usepackage{graphicx}        
\usepackage{multicol}        
\usepackage[bottom]{footmisc}

\makeindex             

\begin{document}

\title*{Shear viscosity of pion gas}
\author{Eiji Nakano}
\institute{Department of Physics and Center for Theoretical Sciences, \\
National Taiwan University, Taipei 10617,Taiwan. \\ 
\texttt{enakano@ntu.edu.tw}}
%
%
\maketitle

\section{Introduction}
\label{sec:1}
Shear viscosity $\eta$ is one of transport coefficients in fluid dynamics, 
which characterizes how viscous the system is in the presence of flow gradient 
\cite{Groot}. 
Since, in general, shear viscosity is inversely proportional to scattering cross-section, 
strongly interacting systems have smaller viscosity 
than weakly interacting ones.  
Recently a universal minimum bound for the ratio of $\eta$ to entropy density $s$ 
has been proposed by Kovtun, Son, and Starinets (KSS) \cite{KOVT1}. 
The bound, $\eta/s \ge \frac{1}{4 \pi}$, is conjectured to be satisfied 
for a large class of strongly interacting quantum field theories 
whose dual descriptions in string theory involve black holes 
in anti-de Sitter space \cite%
{Policastro:2001yc,Policastro:2002se,Herzog:2002fn,Buchel:2003tz}. 
Note that $\eta /s$ is more physical quantity than $\eta$ itself 
because the ratio appears as a diffusion constant of fluid equations. 

In experiments, 
$\eta /s$ close to the minimum bound were found in relativistic 
heavy ion collisions (RHIC) \cite{RHIC,Molnar:2001ux,Teaney:2003pb}. 
This discovery came as a surprise. 
Traditionally, quark gluon plasma (QGP), 
a phase of the quantum chromodynamics (QCD) 
above the deconfinement temperature $T_c \sim 170$ MeV 
at zero baryon density \cite{KL04}, had been thought to be 
in weak interaction regime. 
Partly because lattice QCD simulations of the QGP equation of state above $%
2T_{c}$ are not inconsistent with that of an ideal gas of massless
particles, $e=3p$, where $e$ is the the energy density and $p$ is the
pressure of the system \cite{KL04}. However, recent analyses of the elliptic
flow generated by non-central collisions in RHIC \cite%
{Molnar:2001ux,Teaney:2003pb} and lattice simulations of a gluon plasma \cite%
{Nakamura:2004sy} yielded $\eta /s$ close to the the minimum bound at just
above $T_{c}$. This suggests QGP is strongly interacting at this temperature. 
\footnote{See Ref.~\cite{Asakawa:2006tc} for other possibility to yield 
a small viscosity in expanding QGP.} 
Other implications for strong coupling can be seen 
in sharp peaks of mesonic correlators 
\cite{Hatsuda03,Datta03,Umeda02}, 
and in discussions of the possible microscopic structure of such a state 
\cite{Shuryak:2004tx,Koch:2005vg,Liao:2005pa,GerryEd,GerryRho}. 

Given this situation, 
one naturally gets interested in how $\eta /s$ behaves 
as temperature approaches $T_c$ from below, 
supposing that $\eta /s$ of QCD was already 
close to the minimum bound at just above $T_c$, 
and how it relates to change of the effective degrees of freedom 
through a phase transition or cross over. 

To explore these issues, 
we use chiral perturbation theory (ChPT) and
the linearized Boltzmann equation to perform a model independent calculation
to the $\eta /s$ of QCD in the confinement phase. Earlier attempts to
compute meson matter viscosity using the Boltzmann equation and
phenomenological phase shifts in the context of RHIC hydrodynamical
evolution after freeze out can be found in Refs.~\cite{Davesne,DOBA1,DOBA2}.
In the deconfinement phase, state of the art perturbative QCD calculations
of $\eta $ can be found in Refs.~\cite{Arnold:2000dr,Arnold:2003zc}. 

\subsection{Shear viscosity in Fluid dynamics}
\label{subsec:1-1}

For later understanding it might be worth mentioning 
basic properties of shear viscosity. 
Fluid dynamics describes non-equilibrium system 
where evolution in space and time occurs in macroscopic scale. 
$\eta$ appears as a phenomenological parameter in this scale. 
Usually we can select suitable theory corresponding to 
the characteristic evolution scale of our interest, 
as listed in table~\ref{tab:1}. 

\begin{table}
\centering
\caption{Hierarchy of theories in space-time scales}
\label{tab:1}       
%
%
\begin{tabular}{l|ll}
\hline
Scale & \ Theory  & \\
\hline
Micro 
(range of interaction)
& \ Quantum theory,  &e.g., Linear response theory \ \\ 
Meso  
(mean-free path) 
& \ Kinetic theory, &e.g., Boltzmann eq.  \\ 
Macro 
(sound wave)  
& \ Fluid dynamics, &e.g., Navier-Stokes eq. with $\eta$ \\ 
\hline
\end{tabular}
\end{table}

Fluid dynamics in relativistic framework is consist of two basic equations, 
that is, 
conservation laws of the energy-momentum $T^{\mu\nu}(x)$ 
and the number density (or charge) $n(x)$:  
\begin{eqnarray}
&&\partial_\mu T^{\mu\nu}(x)=0, \\
&&\partial_\mu n(x)V^{\mu}(x)=0,  
\end{eqnarray}
where $V^{\mu}(x)$ is a normalized four velocity of elementary volume, 
$V^{\mu}(x)V_{\mu}(x)=1$. 
In what follows, we will not care the second equation 
in treating pion gas, 
because pion does not carry any conserved charge. 
The energy-momentum tensor in viscous system is decomposed as 
\begin{eqnarray}
T_{\mu\nu}(x)=T_{\mu\nu}^{(0)}(x)+\delta T_{ij}(x), 
\end{eqnarray}
where the first term describes the perfect fluid 
(local equilibrium; frictionless processes) 
with local pressure ${\mathcal{P}}(x)$ and energy density $\epsilon (x)$ 
\begin{eqnarray}
T_{\mu \nu }^{(0)}(x)=\left\{ {
\mathcal{P}}(x)+\epsilon (x)\right\} V_{\mu }(x)V_{\nu }(x)-{\mathcal{P}}%
(x)\delta _{\mu \nu } 
\label{Tzero}, 
\end{eqnarray} 
\footnote{In general, 
in broken phases such as superfluid or chiral condensed phase, 
there might be an another term attributed to the superfluid mode 
independent from the above $T^{(0)}_{\mu\nu}$ in (\ref{Tzero}). 
For details, see Refs.~\cite{Son:1999pa,Pujol:2002na,Lallouet:2002th}. }
, 
and the second term corresponds to a small deviation from local equilibrium, 
which defines shear and bulk viscosities, $\eta$ and $\zeta$, 
\begin{equation}
\delta T_{ij}(x)=-\eta \left( \nabla _{i}V_{j}(x)+\nabla _{j}V_{i}(x)%
-\frac{2}{3}\delta _{ij}\nabla \cdot \mathbf{V}(x)\right) +\zeta \delta
_{ij}\nabla \cdot \mathbf{V}(x). 
\label{TMU}
\end{equation}
One can see that 
$\eta$ ($\zeta$) is in traceless (trace) part of spatial components. 
These transport coefficients, 
which also includes thermal conductivity in the presence of conserved charge, 
are determined by experiments, 
or can be derived from more microscopic theories in principle. 
In general, it becomes more difficult to derive them 
as one begins with smaller scale. 

\begin{figure}
\centering
\includegraphics[height=2.9cm]{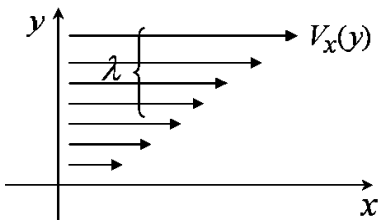}
%
%
\caption{Shear flow on $x$-$y$ plane.}
\label{fig:1}       
\end{figure}

We shall estimate $\eta$ by considering two dimensional case 
depicted in Fig.~\ref{fig:1}, 
where $x$-component of flow velocity $V_x(y)$ varies along $y$ direction.
In other words, shear viscosity is an anisotropic pressure 
defined by the momentum transfer of $x$-component per unit time across unit area 
normal to $y$ direction, in which the force is facing negative $x$ direction.  
Since the momentum transfer occurs within the mean-free path $\lambda$, 
\begin{eqnarray}
\delta T^{\mu\nu}=-\eta \frac{\partial V_x(y)}{\partial y}
\simeq  -n V_y \frac{\partial p_x}{\partial y}\lambda
=-\frac{p_y}{\sigma}\frac{\partial V_x(y)}{\partial y}, 
\end{eqnarray}
where we have used $\lambda=1/(n \sigma)$ 
with $\sigma$ and $n$ being scattering cross-section and density. 
Thus, we can roughly estimate it by dimensional analysis  
up to a numerical factor, 
\begin{eqnarray}
\eta\simeq \frac{T}{\sigma}\simeq \frac{T^3}{|\mathcal{T}|^2}, 
\label{ETAB}
\end{eqnarray}
where we have reduced momentum to temperature $p \sim T$,  
and $\mathcal{T}$ is the scattering amplitude. 
It shows that strong interaction (small $\lambda$) causes 
quick and local equilibrium, 
implying that the system can be well described by hydrodynamics. 
In the case of weak interaction (large $\lambda$), on the other hand, 
a large number of particles (thermal excitations) 
included in $\lambda$ join to relax 
the gradient of flow velocity, so that, it makes $\eta$ finite. 

\subsection{Shear viscosity in Kubo formula}
\label{subsec:1-2}

If one would like to evaluate $\eta$ in field theoretical way with Lagrangian, 
the linear response theory gives a formulation. 
Since $\eta$ is a response of the system 
to the external stimulation $\delta T_{ij}$, 
it is given by Kubo formula 
\begin{eqnarray}
\eta =-\frac{1}{5}\int_{-\infty }^{0}\mathrm{d}t^{\prime }\int_{-\infty
}^{t^{\prime }}\mathrm{d}t\int \mathrm{d}x^{3}\langle \left[
T^{ij}(0),T^{ij}(\mathbf{x},t)\right] \rangle, 
\end{eqnarray}
where $T^{ij}$ the spatial off-diagonal part of energy-momentum tensor, 
and is operator derived from Lagrangian of the system. 
This formula gives the strict definition of $\eta$ 
derived from microscopic Lagrangian. 
One might think a perturbative calculation of the above correlator 
will give an answer for $\eta$. 
But this is not true even for weak coupling theory. 
Indeed, the Kubo formula involves, 
for instance in $\rm{\lambda} \phi^{4}$ theory, 
an infinite number of diagrams at the leading order (LO) \cite{Jeon}. 
This is due to an infrared singularity 
(remember that $\eta$ appears in macroscopic scale). 
However, it is proven that the summation of LO diagrams is equivalent 
to solving the linearized Boltzmann equation (for mesoscopic scale) 
with temperature dependent particle masses and scattering amplitudes \cite{Jeon}. 
The result shows $\eta\sim T^3/\lambda^2$.  
This proof extended the applicable range of the Boltzmann equation to higher 
temperature, but is restricted to weak coupling theories. In the case we are
interested (QCD with $T<140$ MeV), the pion mean free path is always greater
than the range of interaction ($\sim 1$ fm) by a factor of $10^{3}$. Thus,
even though the coupling in ChPT is too strong to use the result of 
$\lambda \phi^{4}$ theory in \cite{Jeon}, 
the temperature is still low enough that 
the use of the Boltzmann equation is justified. 
In the following sections, 
we will show shear viscosity of pion gas 
evaluated from the linearized Boltzmann equation to LO in ChPT, 
and develop arguments on results, 
which is all based on our recent paper Ref.~\cite{JWE1}.

\section{Linearized Boltzmann Equation for Low Energy QCD}

In the hadronic phase of QCD with zero baryon-number density, the dominant
degrees of freedom are the lightest hadrons---the pions. The pion mass $%
m_{\pi }=139$ MeV is much lighter than the mass of the next lightest
hadron---the kaon whose mass is $495$ MeV. Given that $T_{c}$ is only $\sim
170$ MeV, it is sufficient to just consider the pions in the calculation of
thermodynamical quantities and transport coefficients for $T\ll T_{c}$.

The interaction between pions can be described by chiral perturbation theory 
(ChPT) in a systematic expansion in energy and quark ($u$ and $d$
quark) masses \cite{ChPT,GL,Colangelo:2001df}. ChPT is a low energy
effective field theory of QCD. It describes pions as Nambu-Goldstone bosons
of the spontaneously broken chiral symmetry. At $T\ll T_{c}$, the
temperature dependence in $\pi \pi $ scattering can be calculated
systematically. At $T=T_{c}$, however, the theory breaks down due to the
restoration of chiral symmetry.\footnote{%
The QCD chiral restoration happens almost as soon as 
the deconfinement transition at zero baryon density. 
We do not distinguish their critical temperatures.}

Since unlike the Kubo formula the Boltzmann equation
requires semi-classical descriptions of particles with definite position,
energy and momentum except during brief collisions, the mean-free path is
required to be much greater than the range of interaction. Thus the
Boltzmann equation is usually limited to low temperature systems. 
The Boltzmann equation describes the evolution of the isospin averaged pion
distribution function $f=f(\mathbf{x},\mathbf{p},t)\equiv f_{p}(x)$ (a
function of space, time and momentum) as 
\begin{equation}
\frac{p^{\mu }}{E_{p}}\partial _{\mu }f_{p}(x)=\frac{g_{\pi }}{2}%
\int_{123}d\Gamma _{12;3p}\left\{
f_{1}f_{2}(1+f_{3})(1+f_{p})-(1+f_{1})(1+f_{2})f_{3}f_{p}\right\}
\label{BE1}
\end{equation}
where $E_{p}=\sqrt{\mathbf{p}^{2}+m_{\pi }^{2}}$ and $g_{\pi }=3$ is the
degeneracy factor for three pions, 
\begin{equation}
d\Gamma _{12;3p}\equiv \frac{1}{2E_{p}}|\mathcal{T}|^{2}\prod_{i=1}^{3}\frac{%
d^{3}\mathbf{k}_{i}}{(2\pi )^{3}(2E_{i})}\times (2\pi )^{4}\delta
^{4}(k_{1}+k_{2}-k_{3}-p)\ ,
\end{equation}%
and where $\mathcal{T}$ is the scattering amplitude for particles with
momenta $1,2\rightarrow 3,p$. 
LHS of (\ref{BE1}) gives temporal change of $f_{p}(x)$, 
and it should be equal to change by collision (RHS) 
in which the first (second) term corresponds to gain (loss) of 
particles of momentum $p$ by the collision. 

In ChPT, the LO isospin averaged $\pi \pi 
$ scattering amplitude in terms of Mandelstam variables ($s,t$, and $u$) 
is given by 
\begin{equation}
|\mathcal{T}|^{2}=\frac{1}{9}\sum_{I=0,1,2}(2I+1)|\mathcal{T}^{(I)}|^{2}=%
\frac{1}{9f_{\pi }^{4}}\left\{ 21m_{\pi }^{4}+9s^{2}-24m_{\pi
}^{2}s+3(t-u)^{2}\right\} \ .
\end{equation}%
The temperature dependence in pion mass and pion scattering amplitudes can
be treated as higher order corrections.

The distribution function can be decomposed into 
the local equilibrium $f_{p}^{(0)}(x)$ and a deviation from it, 
\begin{equation}
f_{p}(x)=f_{p}^{(0)}(x)+\delta f_{p}(x), 
\end{equation}
where 
$f_{p}^{(0)}(x)=\left( e^{\beta (x)V_{\mu }(x)p^{\mu }}-1\right) ^{-1}$ 
with $\beta (x)$ being the inverse temperature 
and $V^{\mu }(x)$ the four velocity at the space-time point $x$. 
A small deviation $\delta f_{p}(x)$ from the local equilibrium is
parameterized as 
\begin{equation}
f_{p}(x)=f_{p}^{(0)}(x)\left[ 1-\left\{ 1+f_{p}^{(0)}(x)\right\} \chi _{p}(x)%
\right],  
\label{df0}
\end{equation}
while the energy momentum tensor is
\begin{equation}
T_{\mu \nu }(x)
= T_{\mu \nu }^{(0)}(x)+\delta T_{\mu \nu }(x) 
= g_{\pi }\int \frac{\mathrm{d}^{3}\mathbf{p}}{(2\pi )^{3}}%
\frac{p_{\mu }p_{\nu }}{E_{p}}\left[ f_{p}^{(0)}(x)+\delta f_{p}(x)\right].  
\label{dT}
\end{equation}
We will choose the $\mathbf{V}(x)=0$ frame for the point $x$. This implies $%
\partial _{\nu }V^{0}=0$ after taking a derivative on $V_{\mu }(x)V^{\mu
}(x)=1$. Furthermore, the conservation law at equilibrium $\partial _{\mu
}T^{\mu \nu }|_{\chi _{p}=0}=0$ allows us to replace $\partial _{t}\beta (x)$
and $\partial _{t}\mathbf{V}(x)$ by terms proportional to $\nabla \cdot 
\mathbf{V}(x)$ and $\mathbf{\nabla }\beta (x)$. Thus, to the first order in
a derivative expansion, $\chi _{p}(x)$ can be parameterized as 
\begin{equation}
\frac{\chi_{p}(x)}{\beta (x)}=A(p)\nabla \cdot \mathbf{V}(x) 
+B_{ij}(p)\left( \frac{\nabla
_{i}V_{j}(x)+\nabla _{j}V_{i}(x)}{2}
-\delta _{ij}\frac{\nabla \cdot \mathbf{V}(x)}{3}\right),  
\label{df1}
\end{equation}
where $B_{ij}(p) \equiv 
B(p)\left( \hat{p}_{i}\hat{p}_{j}-\frac{1}{3}\delta_{ij}\right)$, 
and $i$ and $j$ are spatial indexes. 
\footnote{
A non-derivative term is not allowed since $f_{p}$ should be reduced to $%
f_{p}^{(0)}$ when $\beta $ and $V^{\mu }$ become independent of $x$. There
is no term with single spatial derivative on $\beta (x)$ either. The only
possible term $\left( \mathbf{V}\cdot \mathbf{\nabla }\right) \beta (x)$
vanishes in the $\mathbf{V}(x)=0$ frame.}

Substituting (\ref{df1}) into the Boltzmann equation, one obtains a
linearized equation for $B(p)$, 
\begin{eqnarray}
\left( p_{i}p_{j}-\frac{1}{3}\delta _{ij}\mathbf{p}^{2}\right) 
&=&\frac{g_{\pi }E_{p}}{2}\int_{123}d\Gamma
_{12;3p}(1+n_{1})(1+n_{2})n_{3}(1+n_{p})^{-1}  \nonumber
\\
&&\times \left[ B_{ij}(p)+B_{ij}(k_{3})-B_{ij}(k_{2})-B_{ij}(k_{1})\right] 
\nonumber \\
&\equiv& g_\pi \hat{F}_{ij}\left[ B\right],  
\label{LB1}  
\end{eqnarray}
where we have dropped the factor $\left( \nabla _{i}V_{j}(x)+\nabla
_{j}V_{i}(x)-\mbox{trace}\right) $ contracting both sides of the equation
and write $f_{i}^{(0)}(x)$ at this point as $n_{i}=\left( e^{\beta
E_{i}}-1\right) ^{-1}$. There is another integral equation associated with $%
\nabla \cdot \mathbf{V}(x)$ which is related to the bulk viscosity $\zeta $
that will not be discussed here. The $\mathbf{\nabla }\cdot \beta $
and $\partial _{t}\mathbf{V}$ terms in $p^{\mu }\partial _{\mu }f_{p}^{(0)}$
will cancel each other by the energy momentum conservation 
in local equilibrium mentioned above.

After putting everything together and 
comparing the energy-momentum tensor (\ref{TMU}) in fluid dynamics 
and (\ref{dT}) in kinetic theory,  we obtain 
\begin{eqnarray}
\eta &=&\frac{g_{\pi }\beta }{10}\int \frac{\mathrm{d}^{3}\mathbf{p}}{(2\pi
)^{3}}\frac{1}{E_{p}}n_{p}\left( 1+n_{p}\right) B_{ij}(p)\left( p_{i}p_{j}-%
\frac{1}{3}\delta _{ij}\mathbf{p}^{2}\right) \   
\\
&=&\frac{g_{\pi }^2 \beta }{10}\int \frac{\mathrm{d}^{3}\mathbf{p}}{(2\pi
)^{3}}\frac{1}{E_{p}}n_{p}\left( 1+n_{p}\right) B_{ij}(p)\hat{F}_{ij}\left[ B%
\right] \equiv g_{\pi}^2 \langle B|\hat{F}[B]\rangle \ .  \label{EQ1}
\end{eqnarray}%
Here one can see immediately that if the scattering cross section is scaled by a
factor $\alpha $, 
\begin{equation}
d\Gamma _{12;3p}\rightarrow \alpha \left( d\Gamma _{12;3p}\right) \ ,
\label{s1}
\end{equation}%
then Eqs.~(\ref{LB1}) and (\ref{EQ1}) imply the following scaling 
\begin{eqnarray}
B_{ij}(p) &\rightarrow &\alpha ^{-1}B_{ij}(p)\ ,  
\\
\eta &\rightarrow &\alpha ^{-1}\eta \ ,  \label{s2}
\end{eqnarray}%
with $\eta $ proportional to the inverse of scattering cross-section. This
non-perturbative result is a general feature for the linearized Boltzmann
equation with two-body elastic scattering.

To find a solution for $B(p)$, one can just solve Eq.~(\ref{LB1}). But here
we follow the approach outlined in Refs.~\cite{DOBA1,DOBA2} to assume that $%
B(p)$ is a\ smooth function which can be expanded using a specific set of
orthogonal polynomials 
\begin{equation}
B(p)=|\mathbf{p}|^{y}\sum_{r=0}^{\infty }b_{r}B^{(r)}(z(p)),
\label{BP1}
\end{equation}%
where $B^{(r)}(z)$ is a polynomial up to $z^{r}$ and $b_{r}$ is its
coefficient. The overall factor $|\mathbf{p}|^{y}$ will be chosen by trial
and error to get the fastest convergence. The orthogonality condition

\begin{equation}
\int \frac{\mathrm{d}^{3}\mathbf{p}}{(2\pi )^{3}}\frac{\mathbf{p}^{2}}{E_{p}}%
n_{p}\left( 1+n_{p}\right) |\mathbf{p}|^{y}B^{(r)}(z)B^{(s)}(z)\propto
\delta _{r,s}\   \label{BP2}
\end{equation}%
can be used to construct the $B^{(r)}(z)$ polynomials up to normalization
constants. For simplicity, we will choose 
$B^{(0)}(z)=1$. 

With this expansion, the consistency condition for $B(p)$ in Eq.~(\ref{EQ1})
yields%
\begin{equation}
\eta =g_\pi \sum_{r}b_{r}L^{(r)}=g_\pi^2 \sum_{r,s}b_{r}\langle B^{(r)}|\hat{F}%
[B^{(s)}]\rangle b_{s}\ ,  \label{ME1}
\end{equation}%
where 
\begin{equation}
L^{(r)}=\frac{\beta }{15}\int \frac{\mathrm{d}^{3}\mathbf{p}}{(2\pi )^{3}}%
\frac{\mathbf{p}^{2}}{E_{p}}n_{p}\left( 1+n_{p}\right) |\mathbf{p}%
|^{y}B^{(r)}(p)\propto \delta _{0,r}. 
\label{LL} 
\end{equation}%
Since $b_{r}$ is a function of $m_{\pi }$, $f_{\pi }$ and $T$, the $b_{r}$'s
in Eq.~(\ref{ME1}) are in general independent functions, such that $%
L^{(r)}=g_\pi \sum_{s}\langle B^{(r)}|\hat{F}[B^{(s)}]\rangle b_{s}$ 
[one can show that this solution of $b_{s}$ gives a unique solution of $\eta $], 
or equivalently
\begin{equation}
\delta _{0,r}L^{(0)}=g_\pi \sum_{s}\langle B^{(r)}|\hat{F}[B^{(s)}]\rangle b_{s}\ .
\label{Lr}
\end{equation}%
This will allow us to solve for the $b_{s}$ and obtain the shear viscosity 
\begin{equation}
\eta =g_\pi b_{0}L^{(0)}.  \label{ETA2}
\end{equation}

In the next section, we will show that this expansion converges well, such
that one does not need to keep many terms on the right hand side of 
Eq.~(\ref{Lr}). If only the $s=0$ term was kept, then 
\begin{equation}
\eta \simeq \frac{\left( L^{(0)}\right) ^{2}}{\langle B^{(0)}|\hat{F}%
[B^{(0)}]\rangle }.
\label{ETA3}
\end{equation}
This resultant formula clearly shows that 
$\eta$ does not depend on the degeneracy factor $g_\pi$. 

The calculation of the entropy density $s$ is more straightforward since $s$%
, unlike $\eta $, does not diverge in a free theory. In ChPT, the
interaction contributions are all higher order in our LO calculation. Thus
we just compute the $s$ for a free pion gas: 
\begin{equation}
s=-g_{\pi }\beta ^{2}\frac{\partial }{\partial \beta }\frac{\log {Z}}{\beta }%
\ ,  \label{S}
\end{equation}%
where the partition function ${Z}$ for free pions is 
\begin{equation}
\frac{\log {Z}}{\beta }=-\frac{1}{\beta }\int \frac{\mathrm{d}^{3}\mathbf{p}%
}{(2\pi )^{3}}\log \left\{ 1-e^{-\beta E(p)}\right\} \ ,
\end{equation}%
up to temperature independent terms. \

\section{Results and Discussion}
\label{sec:3}

\subsection{Shear viscosity of Pion gas}
\label{subsec:3-1}

\begin{figure}
\centering
\includegraphics[height=4.8cm]{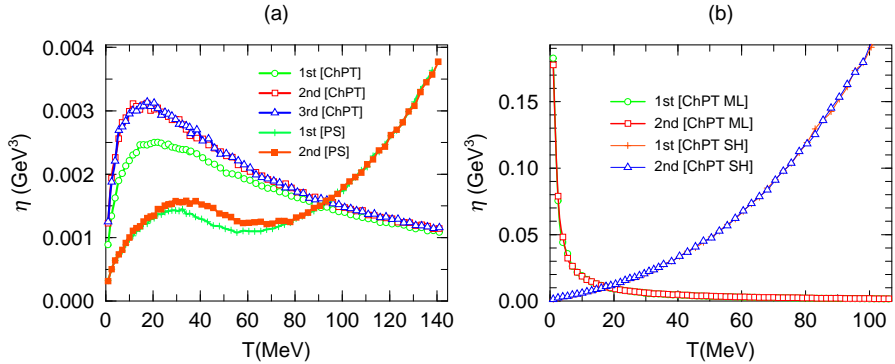}
%
%
\caption{Shear viscosity as a function of temperature. 
(a) for the LO ChPT (ChPT) 
and for an empirical phase-shift scattering amplitude (PS) \cite{Schenk:1991xe}. 
$m_\pi=139$ MeV and $f_\pi=93$ MeV. \ 
(b) for the LO ChPT in the massless limit (ChPT ML) where $f_\pi=87$ MeV, 
and for a constant scattering amplitude in low-energy limit, 
$|\mathcal{T}|^2=\frac{23 m_\pi^4}{9f_\pi^4}$ (ChPT SH). 
$1_{\rm st}$
- $3_{\rm rd}$ show results 
up to the three polynomials in Eq.~(\ref{Lr}).}
\label{fig:2}       
\end{figure}

We present the results for $\eta $ and $\eta /s$ of QCD 
at zero baryon number density. 
Fig.~\ref{fig:2}(a) shows 
the LO ChPT result of $\eta$ using the linearized Boltzmann equation. 
For comparison, we also added 
results for a $\pi$-$\pi$ scattering of empirical phase shifts (PS), 
in which values of parameters were employed from 
Refs.~\cite{Davesne,DOBA1,DOBA2,Schenk:1991xe}. 
Characteristically, 
the PS scattering amplitude does not reflect the phase transition so much 
to be less dependent on momentum than ChPT at high energy region, 
and incorporates resonance effect which is absent in ChPT. 
The lines with circles, squares and triangles correspond to keeping the first
one, two and three polynomials on the right hand side of Eq.~(\ref{Lr}),
respectively. 
We have used $y=0$ and $z(p)=|\mathbf{p}|$ for ChPT and 
$y=2$ and $z(p)=|\mathbf{p}|^2$ for PS to construct the polynomials. 
The figure shows the polynomial expansion converges rapidly. 

As a test of the calculation, 
we also show the shear viscosity result for a constant scattering amplitude 
$|\mathcal{T}|^2=\frac{23 m_\pi^4}{9f_\pi^4}$ in Fig.~\ref{fig:2}(b) (ChPT SH), 
which corresponds to the low energy limit of ChPT and can be mapped onto 
$\lambda \phi^{4}$ theory of Ref.~\cite{Jeon} 
where the scattering amplitude is a constant $|\mathcal{T}|^2=\lambda^2$. 
In $\lambda \phi^{4}$ theory, $\eta $ is monotonically increasing in $T$. 
If $T\gg m_{\phi }$, 
$\eta $ $\propto T^{3}/\lambda ^{2}$ with $T^{3}$ given by dimensional analysis 
and $\lambda^{-2}$ by the scaling of coupling, 
as shown in Eq.~(\ref{ETAB}) and in Eqs.~(\ref{s1}) and (\ref{s2}). 
In ChPT, however, $\eta $ is not monotonic in $T$. At $T\ll m_{\pi }$,
the scattering amplitude is close to a constant, thus ChPT behaves like
a $\lambda \phi ^{4}$ theory. 
But as temperature gets higher to $T\gg m_{\pi }$, 
$\mathcal{T}$ $\propto T^{2}/f_{\pi }^{2}$ 
and thus $\eta $ $\propto f_{\pi }^{4}/T$. 
At what temperature the transition from $\eta $ $\propto T^{3}$ to $\eta $ $\propto
1/T$ takes place depends on the detail of dynamics. In ChPT, this
temperature is around $20$ MeV. 

 It is also interesting to observe the result of ChPT in massless limit, 
Fig.~\ref{fig:2}(b) (ChPT ML). 
$\eta$ blows up at $T=0$. 
It reflects the derivative coupling typical for NG modes,
see Ref.~\cite{Manuel} for $\eta$ of phonon in CFL phase.

\subsection{$\eta/s$ and error estimation}
\label{subsec:3-2}

The radius of convergence in momentum for ChPT is typically 
$4\pi f_{\pi }\sim 1$ GeV. To translate this radius into temperature, 
we compute the averaged center of mass momentum 
$\langle |\mathbf{p}|\rangle =\sqrt{\langle B|\mathbf{p}^{2}|\hat{F}[B]\rangle
/\langle B|\hat{F}[B]\rangle }$. We found that for $T=120$ and $140$ MeV, $%
\langle |\mathbf{p}|\rangle \simeq 460$ and $530$ MeV $<4\pi f_{\pi }$.
However, ChPT is supposed to break down at the chiral restoration 
temperature ($\sim 170$ MeV). Thus our LO\ ChPT result can only be
trusted up to $T\sim 120$ MeV. 
At the next-to-leading order (NLO), 
it is known that the isoscalar $\pi \pi$ scattering length will be increased by 
$\sim 40\%$ \cite{Colangelo:2001df}. 
This will increase the cross section by $\sim 100 \%$ 
and reduce $\eta $ by $\sim 50\%$ near threshold. 
This is an unusually large NLO correction since a typical NLO correction 
at threshold is $\le 20\%$. 
The large chiral corrections does not persist at the higher order. 
At the next-to-next-to-leading order (NNLO), the correction is 
$\sim 10\%$ \cite{Colangelo:2001df}. 
Thus, to compute $\eta $ to $10\%$ accuracy, 
a NLO ChPT calculation is needed.

\begin{figure}
\centering
\includegraphics[height=4.85cm]{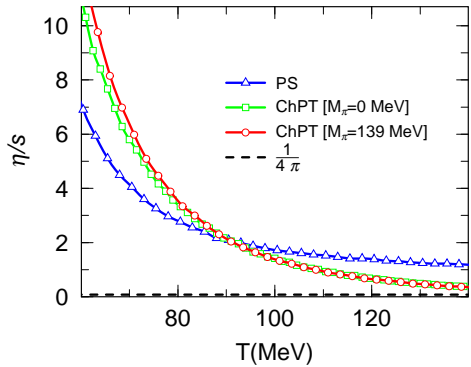}
%
%
\caption{Shear viscosity to entropy density ratios as
functions of temperature. Line with circles (rectangles) is the LO ChPT result 
with $m_{\protect\pi }=139\,(0)$ MeV 
and $f_{\protect\pi}=93(87)$ MeV. Line with triangles is the result using $\protect\pi \protect%
\pi $ phase shifts (PS). Dashed line is the conjectured KSS bound $1/4%
\protect\pi \simeq 0.08$.}
\label{fig:3}       
\end{figure}

The LO ChPT result for $\eta /s$ is shown in Fig.~\ref{fig:3}   
(line with rectangles). 
The error is estimated to be $\sim 50\%$ up to 120 MeV. 
$\eta/s $ is monotonically decreasing and reaches $0.6$ at $T=120$ MeV. 
This is similar to the behavior in the $m_{\pi }=0$ case 
(shown as the line with rectangles) 
where $\eta/s\propto f_{\pi }^{4}/T^{4}$ with $s\propto T^{3}$
from dimensional analysis and $f_{\pi }=87$ MeV in the chiral limit 
\cite{GL,Colangelo:2001df}.
For comparison, 
we also show the result using phenomenological $\pi$-$\pi$
phase shifts \cite{Schenk:1991xe} for $\eta $ but free pions for $s$. 
(Our result for $\eta $ is in good agreement with that of \cite{DOBA2} 
for $T$ between $60$ and $120$ MeV.
 For an earlier calculation using the Chapman-Enskog approximation, 
see Ref.~\cite{Prakash:1993kd}.) 
This amounts to taking into account part of the NLO $\pi \pi$ scattering effects 
but ignoring its temperature dependence and the interaction in $s$. 
Since not all the NLO effects are accounted for, 
this $\eta /s$ is not necessarily more accurate than that using the LO ChPT. 
The comparison, however, gives us some feeling of the size of error 
for the LO result we present here. 
Thus, an error of $\sim 100\%$ at $T=120$ MeV to the LO result 
might be more realistic.

\subsection{Relation between $T_c$ and $\eta/s$}
\label{subsec:3-3}

Naive extrapolations of the three $\eta /s$ curves show that 
the $1/4\pi=0.08$ minimum bound conjectured from string theory 
might never be reached as in phase shift result \cite{dobadoX,NicolaX} 
(the first scenario), or more interestingly, 
be reached at $T\simeq 210$ MeV, as in the LO ChPT result (the second scenario). 
In both scenarios, 
we see no sign of violation of the universal minimum bound for $\eta /s$ below $T_{c}$. \ But to really make sure the
bound is valid from $120$ MeV to $T_{c}$, a lattice computation as was
performed to gluon plasma above $T_{c}$ \cite{Nakamura:2004sy} is needed. In
the second scenario, assuming the bound is valid for QCD, then either a
phase transition or cross over should occur before the minimum bound is
reached at $T\sim 200$ MeV. Also, in this scenario, it seems natural for $%
\eta /s$ to stay close to the minimum bound around $T_{c}$ as was recently
found in heavy ion collisions.

In the second scenario, one might argue that the existence of phase
transition is already known, otherwise we will not have spontaneous symmetry
breaking and the corresponding Nambu-Goldstone boson theory 
at low temperature in the first place. Indeed, it is true in the case of QCD. 
For a spontaneous symmetry breaking theory, the general feature
of $\eta /s$ we see here seems generic. 
\begin{figure}
\centering
\includegraphics[height=3.0cm]{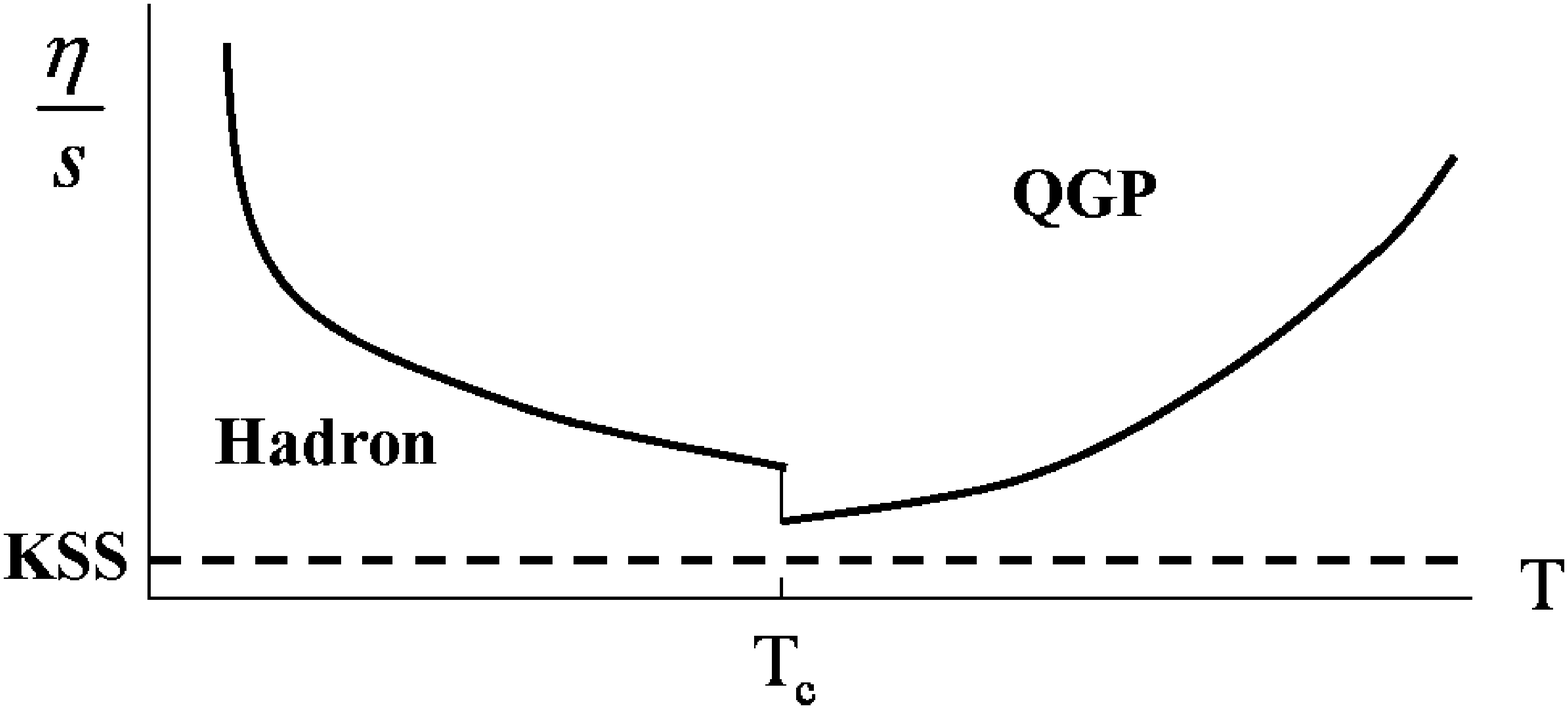}
%
%
\caption{A schematic of $\eta/s$ behavior around the critical temperature.}
\label{fig:4}       
\end{figure}

At asymptotically high $T\gg T_c$ in the deconfinement phase, 
the weak coupling nature of QCD makes $\eta /s$ get higher 
in comparison with that near $T_c$. 
On the other hand, at very low $T\ll T_c$ in the symmetry breaking phase, 
the Nambu-Goldstone bosons are weakly interaction at low temperature, 
thus $\eta /s$ gets lower as $T$ approaches $T_c$ from below. 
A phase transition should occur before the extrapolated $\eta /s$ curve 
coming from high $T$ reaches the bound at $T_{1}$. 
Similarly, a phase transition should occur before the extrapolated $\eta /s$ curve 
coming from low $T$ reaches the bound at $T_{2}$. 
Thus the range of phase transition is $T_{1}\leq T_{c}\leq T_{2}$, 
see Fig.~\ref{fig:4}. 
There might be a discontinuity at $T_c$ 
due to the change of effective degrees of freedom. 
Some relations between $T_{c}$ and the $\eta /s$ bound
are also discussed in Refs.~\cite{Hirano,Csernai:2006zz}.

\subsection{Large $N_f$ and $N_c$ limit}
\label{subsec:3-4}

It is interesting to note that the degeneracy factor $g_{\pi }$ drops out of 
$\eta $ while the entropy $s$ is proportional to $g_{\pi }$, 
as shown in Eqs.~(\ref{Lr}-\ref{ETA3}) and (\ref{S}), respectively. 
This suggests the $\eta /s$ bound might
be violated if a system has a large particle degeneracy factor \cite{KOVT1}. 
For QCD, large $g_{\pi }$ can be obtained
by having a large number of quark flavors $N_{f}$ with $g_{\pi }\sim
N_{f}^{2}$. However, the existence of confinement demands that the number of
colors $N_{c}$ should be of order $N_{f}$ to have a negative QCD beta
function. After using $f_{\pi }\propto \sqrt{N_{c}}$, the combined $N_{c}$
and $N_{f}$ scaling of $\eta /s$ is 
\begin{equation}
\frac{\eta }{s}\propto \frac{f_{\pi }^{4}}{g_{\pi }T^{4}}\propto \frac{%
N_{c}^{2}}{N_{f}^{2}}\ ,
\end{equation}%
which is of order one. Thus QCD with large $N_{c}$ and $N_{f}$ can still be
consistent with the $\eta /s$ bound below $T_{c}$. 



\begin{flushleft} \textbf{\large Acknowledgement} \end{flushleft}
This talk is based on works of close collaboration with Prof. J.-W.~Chen 
at National Taiwan University, and on discussion with 
Prof. T.~Cohen at Maryland University, 
especially on Sec.~\ref{subsec:3-4}. 
The author is supported by the Taiwan's National Science Council 
and in part by National Center for Theoretical Science, 
and also would like to thank organizers and ECT$^*$ 
for arrangement of the talk and local support. 


%
%
%
%



\printindex
\end{document}